\documentclass[11pt]{article}

\usepackage{amssymb}
\usepackage{graphicx}
\usepackage{bm}

\begin{document}


\def \kk{{\bf k}}
\def\qm #1{\ | #1 \rangle}
\def\mq #1{\langle #1 |\ }
\def\norc #1 {\ {\Vert} #1 {\Vert_{c}}}
\def\norN #1 {\ {\Vert} #1 {\Vert_{N}}}
\def\nor #1 {\ {\Vert} #1 {\Vert}}
\def\sous#1#2{\mathop{#1}\limits_{#2}\>}
\def\sur#1#2{\mathop{#1}\limits^{#2}\>}

\def \Ninf#1 {  {\Vert {#1} \Vert_\infty } }
\def \NLip#1 {  {\Vert {#1} \Vert_\theta } }
\def \Norm#1 {  {\Vert {#1} \Vert } }

\def\bZ {{\bf Z}}
\def\bZr {{\bf Z_r}}

\def\kk {{\bf k}}
\def\rr {{\bf r}}

\def \bra{{ \langle}}
\def \ket{{ \rangle}}

\def\cL{{\mathcal{L}}}
\def\cK{{\mathcal{K}}}
\def\cJ{{\mathcal{J}}}
\def\cG{{\mathcal{G}}}
\def\cF{{\mathcal{F}}}
\def\cH{{\cal H}}
\def\cV{{\cal V}}
\def\cU{{\cal U}}
\def\cT{{\cal T}}
\def\cL{{\cal L}}

\renewcommand{\Re}{\mathrm{\,Re}}

\def \lcIN { {\rm in} }

\def \Bra{{ \langle \mskip -3mu \langle}}
\def \Ket{{ \rangle \mskip -3mu \rangle}}
\newtheorem{lemma}{Lemma}

\def \col{ \mskip -2mu : \mskip -2mu}

\def \hik{ \mskip -2mu - \mskip -2mu}
\def \tas{ \mskip -2mu + \mskip -2mu}

\def\bmsig{\mbox{\boldmath $\sigma$}}
\def\bmx{\mbox{\boldmath $x$}}
\def\bmy{\mbox{\boldmath $y$}}

\newcommand{\beq}{\begin{equation}}
\newcommand{\eeq}{\end{equation}}
\newcommand{\barr}{\begin{eqnarray}}
\newcommand{\earr}{\end{eqnarray}}

\newcommand{\andy}[1]{ }

\def\figwidth{8.5cm}


\title{On noise-induced superselection rules}



\title{On noise-induced superselection rules}
\author{S. PASCAZIO\\
\small{Dipartimento di Fisica, Universit\`a di Bari I-70126
  Bari, Italy,}\\
   \small{and Istituto Nazionale di Fisica Nucleare, Sezione di
  Bari, I-70126 Bari, Italy}}
\date{\today}

\maketitle

\begin{abstract}
The dynamical properties of a quantum system can be profoundly
influenced by its environment. Usually, the environment provokes
decoherence and its action on the system can often be schematized
by adding a noise term in the Hamiltonian. However, other
scenarios are possible: we show that by increasing the strength of
the noise, the Hilbert space of the system gradually splits into
invariant subspaces, among which transitions become increasingly
difficult. The phenomenon is equivalent to the formation of the
quantum Zeno subspaces. We explore the possibility that noise can
{\em prevent}, rather than provoke decoherence.
\end{abstract}

\setcounter{equation}{0}
\section{Introduction  }
\label{sec-introd}
\andy{intro}

Interactions with the environment provoke decoherence
\cite{decoherencereview} on quantum systems. The physical
mechanisms at the origin of the loss of quantum coherence are
diverse and can be heuristically modelled in many different ways.
However, usually, these mechanisms can be viewed as yielding a
`disturbance' or a phase randomization of some sort. For this
reason, it is often licit to neglect the detailed features of the
environment and schematize its global effect on the system by
means of noise terms in the Hamiltonian of the latter. One often
reads that noise provokes decoherence. There are, however,
noteworthy exceptions: a large noise can help stabilizing a
quantum system, suppressing transitions to other states. This
mechanism was understood in the late 70's \cite{Harris} and
enabled one to explain the stability of certain chiral molecules.
It is therefore worth investigating in which sense noise can yield
superselection rules and whether/when noise can {\em prevent},
rather than provoke decoherence.

Several strategies have been proposed during the last few years in
order to counter decoherence, in particular in the context of
quantum computation \cite{Review}. Quantum error correcting codes
\cite{ErrorCorrecting}, decoherence-free subspaces
\cite{NoiselessSub}, `bang-bang' pulses and dynamical decoupling
\cite{BBDDsem} are just some examples. Other interesting
proposals make use of the quantum Zeno effect (QZE)
\cite{QZEseminal}) and the recently introduced quantum Zeno
subspaces \cite{FPsuper}. Moreover, the possibility of preserving
quantum coherence by means of a stochastic control has been
recently advocated by Mancini \emph{et al} \cite{Mancini}, who
also emphasized the links with the quantum Zeno effect
\cite{ManciniZeno}.
The unification of these schemes under the same basic ideas
\cite{bang} enables one to look at this problem from a broader
perspective.

In this article we shall look in detail at the afore-mentioned
noise-based strategy to inhibit transitions (and
therefore--perhaps--to control decoherence). We shall start by
looking at a simple example studied by Blanchard, Bolz, Cini, De
Angelis and Serva \cite{bbcds} and Berry \cite{Berry}. We first
reinterpret some of their findings in terms of the QZE
\cite{supercond} and then broaden the applicability of the method
to include a wider class of quantum Zeno phenomena.

\setcounter{equation}{0}
\section{The model}
\label{sec-modello}
\andy{modello}
The model studied by Blanchard \emph{et al} \cite{bbcds} describes
a two-level system interacting with an environment according to
the Hamiltonian
\andy{BBHam}
\beq
H = \alpha \sigma_1 + \beta \eta (t) \sigma_3,
\label{eq:BBHam}
\eeq
where $\alpha$ and $\beta$ are real constants and $\sigma_i \;
(i=1,2,3)$ Pauli matrices. The action of the environment on the
system is modeled by the stochastic term $\eta
\sigma_3$, where
\andy{wn1}
\beq
\langle \eta (t) \rangle = 0, \qquad
\langle \eta (t) \eta(t') \rangle = \delta (t-t'),
\label{eq:wn1}
\eeq
the brackets denoting the average over all possible realizations
of the white noise $\eta$. In terms of the Wiener process
\andy{wiep1}
\barr
& & dW(t)  \equiv W(t+dt)-W(t) = \int_t^{t+dt} \eta(s) ds ,
            \nonumber \\
     & &
\langle dW(t) \rangle = 0, \quad
\langle dW(t) dW(t) \rangle = dt,
\label{eq:wiep1}
\earr
the Ito-Schr\"odinger equation reads $(\hbar=1)$
\andy{strat1}
\beq
| d\psi \rangle = -i \alpha \sigma_1 |\psi \rangle dt
                  -i \beta \sigma_3 |\psi \rangle \circ dW
                = \left( -i \alpha \sigma_1 -\frac{1}{2}\beta^2 \right)
|\psi \rangle dt -i \beta \sigma_3 |\psi \rangle dW,
\label{eq:strat1}
\eeq
where $\circ$ denotes the Stratonovich  product and $|\psi \rangle
= (|\psi_+ \rangle ,|\psi_- \rangle )^T$ is a two-component spinor
(we work in the basis of the eigenstates of $\sigma_3$). When
$\beta=0$, the above equation yields coherent (Rabi) oscillations
between the two eigenstates of $\sigma_3$. This Hamiltonian
schematizes a two-level system interacting with an environment,
whose action is `summarized' by means of a white noise multiplying
an operator of the system. The model describes a superconducting
ring enclosing a quantized magnetic flux. Coherent tunneling
between the two flux configurations is possible if the system is
very well isolated from its environment ($\beta=0$). In general,
coherence is gradually lost when $\beta \neq 0$; however, as we
shall see, it is of primary importance to focus on the timescales
of the decoherence process.

The polarization (Bloch) vector
\andy{polv}
\beq
 \bmx(t)= \langle \psi |\bmsig | \psi \rangle ,
\label{eq:polv}
\eeq
satisfies the stochastic differential equation
\andy{3bbcds}
\beq
  d\bmx(t)=A\bmx(t)dt+B\bmx(t)dW(t),
\label{eq:3bbcds}
\eeq
where
\andy{matricess}
\beq
  A= \left(
    \begin{array}{ccc}
                     -2\beta^2 &   0       &   0          \\
                          0    & -2\beta^2 &   -2\alpha   \\
                          0    & 2\alpha   &   0
    \end{array} \right),
\qquad
  B= \left(
    \begin{array}{ccc}
              0     &-2\beta &   0           \\
             2\beta &   0    &   0          \\
              0     &   0    &   0
    \end{array} \right).
\label{eq:matricess}
\eeq
The Bloch vector is therefore a stochastic process, whose third
component $z=\langle\psi_+|\psi_+ \rangle -
\langle\psi_-|\psi_- \rangle$ yields
information on the probability of finding the system in one of the
eigenstates of $\sigma_3$. The density matrix of a two-level
system (like the one considered above) can always be expressed in
terms of the Bloch vector (\ref{eq:polv}), according to the
formula
\andy{dens2}
\beq
\rho = \frac{1}{2}({\bf 1}+\bmx \cdot \bmsig),
\label{eq:dens2}
\eeq
where $\mbox{Tr}(\rho)=1$ (normalization) and $\mbox{Tr}(\rho
\bmsig) = \bmx$. Pure states are characterized by $\|\bmx\|=1$
and it is easy to check that (\ref{eq:polv}) yields
\andy{3norm}
\beq
\|\bmx(t)\|^2 \equiv x^2(t)+y^2(t)+z^2(t)=1,\qquad\forall t :
\label{eq:3norm}
\eeq
the state remains pure for every individual realization of the
stochastic process. If the average (\ref{eq:wn1})-(\ref{eq:wiep1})
(denoted with a bar throughout) is computed, one gets a
Gorini-Kossakowski-Sudarshan-Lindblad equation \cite{Lindblad}
\andy{Lindbb}
\beq
\frac{d}{dt} \overline{\rho} =
          -i[\alpha \sigma_1,\overline{\rho}] - \beta^2
            (\overline{\rho} - \sigma_3\overline{\rho}\sigma_3).
\label{eq:Lindbb}
\eeq
By making use of the explicit expression (\ref{eq:dens2}) one
obtains
\andy{bbexpl}
\beq
\frac{d}{dt} \overline{x} = -2 \beta^2 \overline{x}, \qquad
\frac{d}{dt} \overline{y} = -2 \alpha \overline{z} -2\beta^2
\overline{y}, \qquad
\frac{d}{dt} \overline{z} = 2 \alpha \overline{y},  \nonumber
\eeq
whose solution is
\andy{bbsol}
\barr
\overline{x}(t) &=& \overline{x}(0) e^{-2 \beta^2t} , \nonumber \\
\overline{y}(t) &=& e^{-\beta^2t} (\overline{y}(0) \cos \omega t
                    +c_1 \sin \omega t) , \label{eq:bbsol} \\
\overline{z}(t) &=& e^{-\beta^2t} (\overline{z}(0) \cos \omega t
                    +c_2 \sin \omega t) ,
\nonumber
\earr
where $c_1=(-\beta^2 \overline{y}(0) -2 \alpha \overline{z}(0)
)/\omega$, $c_2=(\beta^2 \overline{z}(0) +2 \alpha
\overline{y}(0) )/\omega$ and $\omega= \sqrt{4\alpha^2-\beta^4}$.
Note that if $4\alpha^2-\beta^4<0$, $\omega$ becomes purely
imaginary and the solution is simply obtained by replacing the
trigonometric functions in (\ref{eq:bbsol}) with the hyperbolic
ones: $\cos
\omega t
\rightarrow \cosh\omega t,\,
\sin \omega t \rightarrow \sinh\omega t$.

\section{Large noise vs quantum Zeno effect }
\label{sec-QZE}
\andy{QZE}

Different dynamical regimes can be obtained by varying the
coupling $\beta$ with the environment: If $\beta$ is small, the
interaction with the environment is weak and the system undergoes
coherent quantum oscillations between its two states. If, on the
other hand, $\beta$ is large, these oscillations are hindered and
the system becomes `localized' in one of its two states
\cite{bbcds,Berry}.

Let us clarify the links between this localization phenomenon and
the quantum Zeno effect \cite{supercond}. Prepare the system in
the initial state $\overline{x}(0)=
\overline{y}(0) = 0,
\overline{z}(0) =1 $ (all particles in state $|\psi_+ \rangle$).
If the coupling with the environment is large $\beta^2 \gg 2
\alpha$, the solution is
\andy{solz}
\beq
\overline{\bmx}(t) = e^{-\beta^2t}
\left(
    \begin{array}{c}
                     0 \\
          -\frac{2\alpha}{\omega} \sinh \omega t \\
          \cosh \omega t + \frac{\beta^2}{\omega} \sinh \omega t
    \end{array} \right)
\stackrel{ {\rm large} \; \beta^2 }{\longrightarrow}
\left(
    \begin{array}{c}
                     0 \\
          0 \\
          e^{-(2 \alpha^2/\beta^2)t}
    \end{array} \right)
\stackrel{\beta \to \infty}{\longrightarrow}
\left(
    \begin{array}{c}
                     0 \\
          0 \\
          1
    \end{array} \right)
    ,
\label{eq:solz}
\eeq
where `large $\beta^2$' means $\beta^{2} \gg 2\alpha, t^{-1}$ and
we neglected terms $O(\alpha/\beta^2)$ in the third expression. As
one can see, when $\beta$ is large, the oscillations are hindered
and the system tends to remain in its initial state. Notice also
that in the above formulas one implicitly assumes that $t<\infty$.
This `halting' of the quantum evolution due to strong coupling
with the environment is familiar in a variety of physical
situations
\cite{Harris}.

Let us now take a different approach. Assume that the system is
not coupled to the environment $\beta=0$, but {\it frequent}
measurements are performed on the system in order to ascertain
whether it is localized in one of the eigenstates of $\sigma_3$
($|\psi_+ \rangle$ or $|\psi_- \rangle$). This is the usual
framework of `pulsed' observation, typical of the quantum Zeno
effect. The solution of the Bloch equation is (no average is
actually needed, but we keep the bar for ease of comparison with
the previous case)
\andy{solz1}
\beq
\overline{\bmx}(t) =
\left(
    \begin{array}{c}
                     0 \\
          -\sin 2\alpha t \\
          \cos 2\alpha t
    \end{array} \right)
\stackrel{{\rm small}\; t}{\simeq}
\left(
    \begin{array}{c}
                     0 \\
          -2\alpha t   \\
          1 - 2 \alpha^2 t^2
    \end{array} \right) ,
\label{eq:solz1}
\eeq
where `small $t$' means $t \ll 2 \alpha=\omega^{-1}$. It is easy
to check
\cite{supercond} that if $N$ $\sigma_3$-measurements are performed
at time intervals $\delta t$ one gets
\andy{xcollapseN}
\beq
\overline{\bmx}(t) =
\left(
    \begin{array}{c}
                     0 \\
         0 \\
          \left[1-2\alpha^2 \left( \frac{t}{N}\right)^2 \right]^N
    \end{array} \right)
\stackrel{{\rm large}\; N}{\longrightarrow}
\left(
    \begin{array}{c}
                     0 \\
         0 \\
          e^{-(2\alpha^2\delta t)t}
    \end{array} \right)
\stackrel{\delta t\to0}{\longrightarrow}
\left(
    \begin{array}{c}
                     0 \\
         0 \\
          1
    \end{array} \right)
\label{eq:xcollapseN}
\eeq
Notice that we are implicitly assuming that $t<\infty$. Once
again, the oscillations are hindered.

The two situations analyzed in this section, large coupling with
the environment and frequent measurements, yield the same physical
effect. The two regimes can be also \emph{quantitatively}
compared: if
\andy{compar}
\beq
\beta^{-2} = \delta t,
\label{eq:compar}
\eeq
(\ref{eq:solz}) and (\ref{eq:xcollapseN}) are asymptotically
identical. A ($\sigma_3$) white noise of large strength $\beta$
and a series of frequent ($\sigma_3$) observations at short time
intervals $\delta t$ slow down (and eventually halt) the evolution
of an eigenstate of $\sigma_3$ [initial condition $\overline z (0)
= 1$].

\section{The general framework }
\label{sec-genfram}
\andy{genfram}
We can now generalize the results of the previous sections in
order to try and understand the reasons of the occurrence of the
`localization' phenomenon in the initial state (which was also an
eigenstate of $\sigma_3$). Since a large noise is physically
equivalent to the quantum Zeno effect and since the latter is
physically equivalent to dynamical decoupling \cite{bang} and
leads to the formation of the quantum Zeno subspaces
\cite{FPsuper}, one expects that the `localization' observed in
the preceding sections can be viewed as a \emph{dynamical}
phenomenon, due to the formation of a Zeno subspace. This
expectation is correct and can be put on firm ground.

Let a quantum system be described by the time-dependent
Hamiltonian
\andy{iHam}
\beq
H_K = H_0 + \eta (t) K H_1,
\label{eq:iHam}
\eeq
where $H_0$ and $H_1$ are Hermitian, time-independent operators.
The action of the environment on the system is schematized by the
stochastic term $\eta K H_1$, where $\eta$ is a white noise and
$K$ the coupling constant. The Hamiltonian (\ref{eq:BBHam}) is a
particular case of the above.

The evolution is
\andy{strat}
\beq
|d\psi\rangle=-iH_0|\psi\rangle dt
                    -i K H_1 |\psi \rangle \circ dW
             =\left(-iH_0-\frac{1}{2} K^2 H_1^2\right)|\psi\rangle dt
                    -i K H_1 |\psi \rangle dW,
\label{eq:strat}
\eeq
or alternatively
\andy{Lindy}
\beq
\frac{d}{dt} \overline{\rho} =
          -i[H_0,\overline{\rho}] - \frac{K^2}{2}
                    \{ H_1^2,\overline{\rho} \} +
                    K^2 H_1 \overline{\rho} H_1 =
(\cL_0 + K^2 \cL) \overline{\rho}.
\label{eq:Lindy}
\eeq
where $[\cdot,\cdot]$ is the commutator, $\{\cdot,\cdot
\}$ the anticommutator and $\cL_0$ and $\cL$ are the free and
dissipative part of the Liouvillian, respectively.

Let us endeavor to understand what happens when $K$ becomes large.
Consider the limiting evolution operator in the interaction
picture
\beq
\label{eq:limevol}
\cU(t)=\lim_{K\to\infty}U_K^{\rm
I}(t)=\lim_{K\to\infty}U_1^\dagger (t)U_K (t),
\eeq
where
\barr
& & U_K(t)= \exp (-i H_K t), \nonumber \\
& & U_1(t) =
\exp\left(-i KH_1\int_0^t\eta(t')dt'\right)
= \exp\left(-i KH_1W(t)\right) ,
\label{eq:U1inter}
\earr
all evolution operators acting \emph{\`{a} la} Ito on the wave
function. $U_K^{\rm I}$ satisfies the Schr\"odinger equation in
the interaction picture
\beq
i \partial_t U_K^{{\rm I}} (t) = H^{{\rm I}}_0(t) U_K^{{\rm I}}
(t) , \qquad H^{{\rm I}}_0(t)=U_1(t)^\dagger H_0 U_1(t)
\eeq
and it is not difficult to show, by adapting the proof of Ref.\
\cite{FPsuper}, that in the large-$K$ limit the evolution operator
becomes diagonal with respect to $H_1$:
\beq
\label{eq:diagevol}
[\cU(t), P_n]=0, \quad{\rm where}\quad H_1 P_n=\eta_n P_n,
\eeq
$P_n$ being the orthogonal projection onto $\cH_{P_n}$, the
eigenspace of $H_1$ belonging to the eigenvalue $\eta_n$. [Note
that in Eq.\ (\ref{eq:diagevol}) the eigenvalues are in general
distinct, $\eta_n\neq\eta_m$ for $n\neq m$, and the $\cH_{P_n}$'s
are in general multidimensional.] Moreover, the limiting evolution
operator has the explicit form
\beq
\label{eq:theorem}
\cU(t)=\exp(-iH_{{\rm diag}} t),\qquad H_{{\rm diag}}=\sum_n P_n
H_0 P_n \equiv \hat P H_0.
\eeq
In words, in the $K\to\infty$ limit an effective superselection
rule arises and the total Hilbert space is split into (Zeno)
subspaces $\cH_{P_n}$ that are invariant under the evolution. The
dynamics within each Zeno subspace $\cH_{P_n}$ is governed by the
diagonal part $P_n H_0 P_n$ of the free Hamiltonian $H_0$. We
stress that the superselection rules discussed here are a
consequence of the Zeno dynamics (strong coupling) and are
equivalent to the celebrated `W$^3$' ones \cite{WWW}.

We also notice that the very same Zeno subspaces could be obtained
by looking for the eigenspace of the dissipative part of the
Liouvillian $\cL$ in (\ref{eq:Lindy}) corresponding to the null
eigenvalue:
\beq
\cL \hat P =0 ,
\eeq
where $\hat P$ is defined in (\ref{eq:theorem}). Since a vanishing
eigenvalue implies in this case no dissipation, the corresponding
Zeno subspaces can be viewed as decoherence-free.

\section{Comments}
\label{sec-concom}
\andy{concom}

We have analyzed a method to inhibit quantum transitions that
makes use of a large noise. The method is well known since long
ago \cite{Harris,bbcds,Berry}, but the interpretation in terms of
the quantum Zeno subspaces \cite{FPsuper} is novel. A complete
theory, valid for general Gorini-Kossakowski-Sudarshan-Lindblad
equations \cite{Lindblad} will be presented elsewhere, as it is
far from being trivial. Such a complete theory would be required,
in particular, in order to fully understand some recent proposals
\cite{Mancini,ManciniZeno} that focus on the preservation of
quantum coherence by stochastic control. The real problem, when
one endeavors to control decoherence \cite{Calarco} is the
occurrence of the \emph{inverse} Zeno effect \cite{IZE} and the
key role played by the form factors of the interaction. In order
to take the consequences of the inverse Zeno effect into account
it is important to accurately model the interaction between the
quantum system and its environment. It is well known that there is
no general recipe in order to get `noise' terms from the total
Hamiltonian (describing the environment + the system) in a
rigorous way. As a matter of fact, this program can be carried out
only in some particular cases
\cite{FKM}, that have played a fundamental role in
clarifying the features of quantum dissipative phenomena
\cite{Wax}. However, strong coupling regimes should be handled
separately and the validity of the interaction Hamiltonian in
(\ref{eq:iHam}), when one endeavors to model the physical system
of interest, must be carefully pondered over. Other issues that
are certainly worth exploring, in this context, are the links with
the so-called continuous measurements
\cite{varicont} and the mechanisms yielding stochastic resonance
\cite{stochres}.

\section*{Acknowledgements}

I thank my friends Paolo Facchi and Hiromichi Nakazato for many
suggestions and clarifications. The proof sketched in Sec.\
\ref{sec-genfram} was suggested by Paolo Facchi; mine was more
complicated.


\end{document}